\documentclass[conference]{IEEEtran}
\IEEEoverridecommandlockouts

\usepackage{tikz}
\usetikzlibrary{arrows.meta,
                chains,
                positioning,
                shapes.geometric,
                shapes.symbols,
                decorations.pathreplacing,
                decorations.pathmorphing,
                calligraphy
                }

\usepackage{cite}
\usepackage{amsmath,amssymb,amsfonts,amsthm}
\usepackage{algpseudocode, algorithm}
\usepackage{tabularx,colortbl}
\usepackage{adjustbox}
\usepackage{comment}
\usepackage{pifont}
\usepackage{graphicx}
\usepackage{textcomp}
\usepackage{booktabs}
\usepackage{diagbox}
\usepackage{caption}
\usepackage{subcaption}
\usepackage{multirow, multicol}
\usepackage{soul}
\usepackage{xspace}
\usepackage{MnSymbol,bbding,pifont}
\usepackage{hyperref}
\usepackage{enumitem}

\usetikzlibrary{automata,arrows,positioning,shapes.geometric,decorations.pathmorphing}
\usetikzlibrary{arrows.meta}

\usepackage{balance}

\newcommand{\RN}[1]{%
  \textup{\uppercase\expandafter{\romannumeral#1}}%
}

\usepackage{fontawesome}
\usepackage{xcolor}
\usepackage{array,colortbl,xcolor}
\usepackage{color}

\def\BibTeX{{\rm B\kern-.05em{\sc i\kern-.025em b}\kern-.08em
    T\kern-.1667em\lower.7ex\hbox{E}\kern-.125emX}}
    
\usepackage{tcolorbox}

\definecolor{ao}{rgb}{0.12, 0.3, 0.17}
\definecolor{darklavender}{rgb}{0.45, 0.31, 0.59}

\newcommand{\eclipsecolor}[1]{\textcolor{eclipse}{#1}}

\newcommand{\boldparagraph}[1]{\vskip 0.05in\noindent\textbf{#1.}}

\newcommand{\tool}{{\sffamily LEGOS-SLEEC}\xspace}
\newcommand{\sota}{{\sffamily LEGOS-SLEEC}\xspace}
\newcommand{\tfol}{FOL$^*$\xspace}

\definecolor{turquoisegreen}{rgb}{0.63, 0.84, 0.71}
\definecolor{tuftsblue}{rgb}{0.28, 0.57, 0.81}
\definecolor{atomictangerine}{rgb}{1.0, 0.6, 0.4}
\definecolor{lightapricot}{rgb}{0.99, 0.84, 0.69}
\definecolor{mygray}{rgb}{0.5,0.5,0.5}
\definecolor{mymauve}{rgb}{0.1,0.2,0.7}
\definecolor{olivegreen}{cmyk}{.6,.4,0.8,0}

\definecolor{green}{rgb}{0.0, 0.65, 0.31}
\definecolor{jasper}{rgb}{0.84, 0.23, 0.24}
\definecolor{redniki}{rgb}{0.99, 0.40, 0.40}
\definecolor{cookie}{rgb}{1.0, 0.86, 0.35}
\definecolor{cottoncandy}{rgb}{1.0, 0.74, 0.85}
\definecolor{lightapricot}{rgb}{0.99, 0.84, 0.69}
\definecolor{eclipse}{RGB}{127,0,85}

\sethlcolor{lightapricot}

\makeatletter
\algrenewcommand\ALG@beginalgorithmic{\scriptsize}
\makeatother

\theoremstyle{definition}

\newcommand{\sleec}[1]{{\small\fontfamily{qcr}\selectfont#1}}
\newcommand{\dsl}{\texttt{\small SLEEC DSL}\xspace}
\newcommand{\sleeckeyword}[1]{\eclipsecolor{\textbf{#1}}}

\def\BibTeX{{\rm B\kern-.05em{\sc i\kern-.025em b}\kern-.08em
    T\kern-.1667em\lower.7ex\hbox{E}\kern-.125emX}}
\begin{document}

\title{LEGOS-SLEEC: Tool for Formalizing and Analyzing Normative Requirements}

\author{\IEEEauthorblockN{Kevin Kolyakov, Lina Marsso, Nick Feng, Junwei Quan, Marsha Chechik}
\IEEEauthorblockA{
\textit{University of Toronto,}
Toronto, Canada \\
\{kolyakov,lmarsso,fengnick,jwquan,chechik\}@cs.toronto.edu}
}

\maketitle

\begin{abstract}
Systems interacting with humans, such as assistive robots or chatbots, are increasingly integrated into our society. To prevent these systems from causing social, legal, ethical, empathetic, or cultural (SLEEC) harms, normative requirements specify the permissible range of their behaviors. These requirements encompass both functional and non-functional aspects and are defined with respect to time. Typically, these requirements are specified by stakeholders from a broad range of fields, such as lawyers, ethicists, or philosophers, who may lack technical expertise.  Because such stakeholders often have different goals, responsibilities, and objectives, ensuring that these requirements are well-formed is crucial. \sleec{SLEEC} DSL, a domain-specific language resembling natural language, has been developed to formalize these requirements as \sleec{SLEEC} rules. In this paper, we present {\sffamily LEGOS-SLEEC}, a tool designed to support interdisciplinary stakeholders in specifying normative requirements as \sleec{SLEEC} rules, and in analyzing and debugging their well-formedness. {\sffamily LEGOS-SLEEC} is built using four previously published components, which have been shown to be effective and usable across nine case studies. Reflecting on this experience,  we have significantly improved the user interface of {\sffamily LEGOS-SLEEC} and its diagnostic support, and demonstrate the effectiveness of these improvements using four interdisciplinary stakeholders. Showcase video URL is \url{https://youtu.be/LLaBLGxSi8A}
\end{abstract}


\section{Introduction}
\label{sec:intro}
Nowadays, an increasing number of systems interact with our society and environments, such as assistive robots, chatbots, or forest protection drones. It is important to specify \textit{normative requirements} to prevent these systems from causing social, legal, ethical, empathetic, or cultural (SLEEC) harms. For example, a legal requirement for an assistive robot might state: ``Ra: When the user tells the robot to open the curtains then the robot should open the curtains, unless the user is ‘undressed’''\cite{daisy-22}.
Normative requirements encompass both functional and non-functional aspects and are defined with respect to time, and are elicited from a diverse group of interdisciplinary stakeholders (e.g., lawyers and ethicists) who may lack technical expertise and have differing goals and priorities As a result, articulating these requirements poses challenges, making it important to support  interdisciplinary stakeholders in ensuring that the elicited requirements do not have \textit{well-formedness issues} (WFIs)~\cite{Getir-Yaman-et-al-23,feng-et-al-23-b,feng-et-al-24} such as conflicts~\cite{Getir-Yaman-et-al-23}.

Previous works have proposed a domain-specific language (DSL), \sleec{SLEEC DSL}, to formalize these requirements as \sleec{SLEEC} rules~\cite{Getir-Yaman-et-al-23}. \sleec{SLEEC DSL} has been shown to be accessible to stakeholders from different fields, including lawyers, philosophers, psychologists, ethics, doctors, roboticists, and software engineers~\cite{Getir-Yaman-et-al-23,feng-et-al-23-b,feng-et-al-24,feng-et-al-24bb}. For instance, the requirement Ra is expressed in \sleec{SLEEC DSL}  rule \sleec{r2} in Tbl.~\ref{tab:sleecRules}. 

Our goal is to help a broad range of stakeholders elicit a coherent and well-formed set of normative requirements. To this end, we previously proposed approaches~\cite{feng-et-al-23-b,feng-et-al-24} to verify whether normative requirements expressed as \sleec{SLEEC} rules are conflicting, redundant, sufficient to prevent undesirable behavior, or overly restrictive —-- issues collectively referred to as \emph{well-formedness constraints (WFIs)}. We also developed an approach~\cite{feng-et-al-24c} that generates WFI verification diagnostics accessible to interdisciplinary stakeholders, along with a methodology~\cite{feng-et-al-24bb} to integrate these components into the elicitation and validation process for normative requirements.

In this paper, we present \tool, an integrated development environment (IDE) designed to support interdisciplinary stakeholders in specifying normative requirements as \sleec{SLEEC} rules and analyzing and debugging their well-formedness. \tool integrates four previously published components~\cite{feng-et-al-23,feng-et-al-23-b,feng-et-al-24,feng-et-al-24c} that have been shown to successfully assist interdisciplinary stakeholders across several case studies, including  assistive-care robots, tree-disease detection drones, and collaborative robots in manufacturing. Drawing from our experience, we have significantly improved the user interface and enhanced the diagnostic support in \tool by removing irrelevant information and incorporating essential missing details to better support stakeholders in debugging WFIs.

The main contributions of our paper are:
(1) A preliminary study identifying challenges in eliciting and analyzing normative requirements using a previous version of \tool; 
(2) \tool, an IDE that provides support for the \sleec{SLEEC} rules elicitation process and improves diagnostic feedback for WFIs;
(3) An evaluation involving four multidisciplinary stakeholders across a range of real-world case studies, demonstrating the effectiveness and usability of \tool; and
(4) A short video showcasing our tool.

The rest of this paper is organized as follows.
Sec.~\ref{sec:background} gives the background material for our work.
Sec.~\ref{sec:tool} present our \tool tool. 
Sec.~\ref{sec:evaluation} presents the evaluation of the \tool usability, and we conclude in Sec.~\ref{sec:conclusion}.

\section{Background}
\label{sec:background}
In this section, we provide an overview of \sleec{SLEEC} DSL, and the five \sleec{SLEEC}  well-formedness properties checked by \tool. 

\boldparagraph{SLEEC DSL}
The key concepts of the \sleec{SLEEC} DSL~\cite{Getir-Yaman-et-al-23} consist of definitions and rules (see Tbl.~\ref{tab:sleecRules}).
Definitions declare events and measures representing system capabilities and its activities during the interaction with the environment, including humans. 
\sleeckeyword{Events} represent instantaneous actions whereas \sleeckeyword{measures} represent capabilities to provide (immediately) information captured by values of data types, such as \sleeckeyword{Boolean}, \sleeckeyword{numeric}, and \sleeckeyword{scale}. In the SLEEC DSL, events are capitalized while measures are not. 
Rules have the basic form ``\sleec{\sleeckeyword{when} trigger \sleeckeyword{then} response}''. Such a rule defines the required response when the event in the trigger happens and its conditions on measures, if any, are satisfied. For example, rule \sleec{r2} applies when the event \sleec{SmokeDetectorAlarm} occurs, in which case the response \sleec{CallEmergencyServices} is required within the next $300$ seconds. 
A \dsl rule can be accompanied by one or more \emph{defeaters}, introduced using the ``\sleec{\sleeckeyword{unless}}'' construct. The language incorporates time constructs allowing responses with deadlines and timeouts using the ``\sleec{\sleeckeyword{within}}'' construct, as seen in rule \sleec{r2}. 

\begin{table*}[t]
    \centering
    \caption{{\small Normative requirements for the assistive robot expressed in \dsl.}}
    \label{tab:sleecRules}

    \vspace{-0.1in}
    \scalebox{0.71}{
        \begin{tabular}{r  l}
        \\
        \toprule
        \multicolumn{2}{c}{\sleec{\textbf{Definitions}}}\\
        \midrule
             \sleec{\sleeckeyword{event}} & \sleec{HumanOnFloor}\\
             \sleec{\sleeckeyword{event}} & \sleec{CallEmergencyServices}\\
             \sleec{\sleeckeyword{event}} & \sleec{SmokeDetecorAlarm}\\
             \sleec{\sleeckeyword{event}} & \sleec{FireSafetyMeasures}\\
             \sleec{\sleeckeyword{measure}} & \sleec{humanAssents: \sleeckeyword{boolean}} \\
             \sleec{\sleeckeyword{measure}} & \sleec{userDisablesAlarm: \sleeckeyword{boolean}} \\
             \sleec{\sleeckeyword{measure}} & \sleec{userUnconscious: \sleeckeyword{boolean}} \\
             \sleec{\sleeckeyword{measure}} & \sleec{userDistressed: \sleeckeyword{scale}(low, medium, high)} \\
             \sleec{\sleeckeyword{constant}} & \sleec{MIN\_TEMP = 16}\\
        \bottomrule
    \end{tabular}
    }
    \scalebox{0.71}{
        \begin{tabular}{rl}
        \\
        \toprule
        \multicolumn{2}{c}{\sleec{\textbf{Rules}}}\\
        \midrule
             \sleec{r1:=} & \sleec{\sleeckeyword{when} HumanOnFloor \sleeckeyword{and} (\sleeckeyword{not} humanAssents) \sleeckeyword{then}} \\
             & \sleec{\sleeckeyword{not} CallEmergencyServices  \sleeckeyword{within} 600 \sleeckeyword{seconds}} \\
             \sleec{r2:=} &  \sleec{\sleeckeyword{when} OpenCurtainRequest \sleeckeyword{and} (\sleeckeyword{not} underDressed) \sleeckeyword{then}} \\
             & \sleec{OpenCurtain \sleeckeyword{within} 30 \sleeckeyword{seconds}} \\
             \sleec{r3:=} & \sleec{\sleeckeyword{when} SmokeDetecorAlarm \sleeckeyword{then} CallEmergencyServices \sleeckeyword{within} 300 \sleeckeyword{seconds}} \\
             \sleec{r4:=} &  \sleec{\sleeckeyword{when} DressingStarted \sleeckeyword{and}({roomTemperature} $<$ MIN\_TEMP)} \\
             & \sleec{\sleeckeyword{and} {userUnderDressed}) \sleeckeyword{then} DressingComplete  \sleeckeyword{within} 1 \sleeckeyword{minutes}}\\
             \sleec{c1:=} &  \sleec{\sleeckeyword{when} SmokeDetecorAlarm \sleeckeyword{and}((\sleeckeyword{not} userDisablesAlarm) \sleeckeyword{or} alarmRestarts)} \\
             & \sleec{\sleeckeyword{then} \sleeckeyword{not} CallEmergencyServices  \sleeckeyword{within} 1 \sleeckeyword{minutes}}\\
        \bottomrule
        \end{tabular}
    }
    \vspace{-0.2in}
\end{table*}

\boldparagraph{Well-formedness properties}
Below, we briefly review five \sleec{SLEEC}  well-formedness properties~\cite{feng-et-al-24}: vacuous- and situational-conflicts, redundancies, restrictiveness, and insufficiency, focusing on situational conflicts and insufficiency (bold) since they are used in Sec.~\ref{sec:tool}. 
\textit{(1) Vacuous Conflicts:}
A rule is \emph{vacuously conflicting} if the trigger of the rule is not in the accepted behaviours defined by the rule set, i.e., triggering the rule in any situation will lead to a violation of some rules and thus cause a conflict. 
\textbf{\textit{(2) Situational Conflicts:}}
A rule $r$ is situationally conflicting if there exists a feasible $r$-triggering situation that eventually causes a conflict. 
{For instance, consider the \sleec{SLEEC} rules \sleec{r1} and \sleec{r3} shown in Tbl.~\ref{tab:sleecRules}. \sleec{r1} conflicts with \sleec{r3} in a situation where the event \sleec{HumanOnFloor} occurs without human consent for calling emergency services (\sleec{\sleeckeyword{not} humanAssents}), while the event \sleec{SmokeDetectorAlarm} is also triggered. This results in \sleec{r3} blocking the response of \sleec{r1}, as illustrated at the top of Fig.~\ref{fig:overExample}.} 
\textit{(3) Redundancy:}
A rule in a rule set  is \emph{redundant} if removing it does not increase the behaviours allowed by the rule set. 
\textit{(4) Restrictive Rules:}
A rule set is overly restrictive if it is impossible to execute the desirable behavior while respecting every rule. 
\textbf{\textit{(5) Insufficient Rules:}}
A set of rules are insufficient if  adhering to those rules can still allow executing an  undesirable behavior.  
{For instance, consider the concern \sleec{c1} shown in Tbl.~\ref{tab:sleecRules}. \sleec{c1} can arise even when all the \sleec{SLEEC} rules in Tbl.~\ref{tab:sleecRules} are respected, as illustrated in the situation described at the top of Fig.~\ref{fig:overUnderExample}.}

\section{LEGOS-SLEEC Tool}
\label{sec:tool}
In this section, we describe the architecture of \tool and briefly  present the preliminary study conducted with eight stakeholders to understand the effectiveness of the tool and  collect feedback on required improvements to better support them in  eliciting well-formed \sleec{SLEEC} rules. We then describe two new components in more detail and finish with \tool implementation details.  

\subsection{Architecture}
\tikzset{
    tool/.style={draw,rectangle,rounded corners,minimum width=1.8cm,minimum height=0.9cm,align=center,text width=1.5cm,fill=black!20,font=\small\sffamily},
    toold/.style={draw=red,line width=2pt, dotted,rectangle,rounded corners,minimum width=2cm,minimum height=0.9cm,align=center,text width=1cm,fill=red!20,font=\small\sffamily},
    tool-n/.style={draw,rectangle,rounded corners,minimum width=2cm,minimum height=0.9cm,align=center,text width=2cm,fill=white,font=\small\sffamily},
    tool_carla/.style={draw,rectangle,rounded corners,minimum width=1.4cm,minimum height=0.9cm,align=center,text width=1.5cm,fill=white,font=\small\sffamily},
    tool_carlaLong/.style={draw,rectangle,rounded corners,minimum width=2.9cm,minimum height=0.9cm,align=center,text width=2.8cm,fill=white,font=\small\sffamily},
    tool_carlaLongL/.style={draw,rectangle,rounded corners,minimum width=3.2cm,minimum height=0.9cm,align=center,text width=3.8cm,fill=white,font=\small\sffamily},
    artifact/.style={draw=white,draw opacity=0.3,trapezium,trapezium left angle=82,trapezium right angle=98,text width=2.5cm,align=center,font=\small\sffamily},
    artifactni/.style={draw=white,draw opacity=0.3,trapezium,trapezium left angle=82,trapezium right angle=98,text width=5cm,align=center,font=\small\sffamily},
    artifactlina/.style={draw=white,draw opacity=0.3,trapezium,trapezium left angle=82,trapezium right angle=98,text width=5cm,align=center,font=\small\sffamily},
    artifactd/.style={draw=white,trapezium,trapezium left angle=82,trapezium right angle=98,text width=2.8cm,align=center,font=\small\sffamily},
    artifactmm/.style={draw=white,trapezium,trapezium left angle=82,trapezium right angle=98,text width=2.6cm,align=center,font=\small\sffamily},
    artifactbb/.style={draw=white,trapezium,trapezium left angle=82,trapezium right angle=98,text width=2cm,align=center,font=\small\sffamily},
    artifactbbb/.style={draw=white,rectangle,text width=2.6cm,align=center,font=\small\sffamily},
    artifact-b/.style={draw=white,trapezium,trapezium left angle=82,trapezium right angle=98,text width=2.8cm,align=center,font=\small\sffamily},
    artifact-s/.style={draw=white,trapezium,trapezium left angle=82,trapezium right angle=98,text width=1.6cm,align=center,font=\small\sffamily},
    progl/.style={draw,tape,tape bend top=none,align=center,text width=2cm,fill=gray!20,font=\small\sffamily}
}
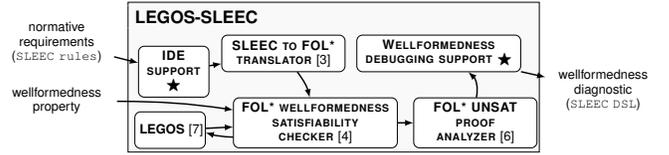
\begin{figure}[t]
    \centering
\scalebox{.54}{
\begin{tikzpicture}[x=2.25cm,y=0.9cm]
\draw[fill=lightgray!10] (0.6,6.2) rectangle (5.1,2.1);


\node[artifactni] (name) at (1.35,5.8) {{\large \textbf{\tool}}};
\node[artifact] (req) at (-0.18,5.1) {normative requirements (\sleec{SLEEC rules})};
\node[artifactmm, below = 10 pt of req] (prop) {wellformedness property};
\node[tool_carla] (IDE) at (1.1,4.3) {\textbf{\textsc{IDE support} $\bigstar$}};
\node[tool_carlaLong] (trans) at (2.3,4.8) {\textbf{\textsc{SLEEC to FOL* translator}}~\cite{feng-et-al-23-b}};
\node[tool_carlaLongL] (wfiChecker) at (2.65,2.9) {\textbf{\textsc{FOL* wellformedness \\ satisfiability checker}}~\cite{feng-et-al-24}};
\node[tool_carlaLong] (proof) at (4.4,2.9) {\textbf{\textsc{FOL* UNSAT proof analyzer}}~\cite{feng-et-al-24c}};
\node[tool_carla] (smt) at (1.06,2.7) {\textbf{{\sffamily LEGOS}}~\cite{feng-et-al-23}};
\node[artifactbbb] (out1) at (5.8,3.8) {wellformedness diagnostic \\ (\sleec{SLEEC DSL})};
\node[tool_carlaLongL] (check) at (4,4.8) {\textbf{\textsc{Wellformedness debugging support} $\bigstar$}};


\begin{scope}[every path/.style={-latex}]
\draw [line width=1pt] 
        (req) edge (IDE)
       (IDE) edge (trans)
       (trans) edge [bend left = 12](wfiChecker)
       (prop) edge [bend left=4] (wfiChecker)
       (wfiChecker) edge [bend left=7] (smt)
       (smt) edge (wfiChecker)
       (wfiChecker) edge (proof)
       (proof) edge [bend right] (check)
       (check) edge (out1)
      ;
      
       

\end{scope}
\end{tikzpicture}
}

\caption{\small Architecture of \tool ($\bigstar$ - {new components}).}
\label{fig:architecture-c}
\vspace{-0.15in}
\end{figure}
\tool provides an integrated development environment (IDE) where stakeholders can: (1) elicit normative requirements as \sleec{SLEEC} rules (e.g., Tbl.~\ref{tab:sleecRules}), and (2) select one of five types of well-formedness properties to check—vacuous conflict, situational conflict, insufficiency, overly restrictive rules, and redundancy (see Sec.~\ref{sec:background}). When a property is selected, \tool generates a diagnosis, highlighting the cause of any well-formedness issue detected.  The architecture of \tool is 
shown in Fig.~\ref{fig:architecture-c}. The components marked with a star ({\small$\bigstar$}) have been newly introduced and will be discussed in more detail below.

Once the \sleec{SLEEC} rules are elicited, stakeholders can select the well-formedness properties to check. The rules are then translated into First-Order Logic with quantifiers over relational objects (FOL$^*$) by the \textsc{SLEEC to FOL$^*$ translator}~\cite{feng-et-al-23-b}. This component calls the \textsc{FOL$^*$ well-formedness satisfiability checker}~\cite{feng-et-al-24}, which converts the well-formedness of \sleec{SLEEC} rules into an FOL$^*$ satisfiability problem and invokes the FOL$^*$ Satisfiability checker \textsc{LEGOS}~\cite{feng-et-al-23} to solve it.

The \textsc{FOL$^*$ well-formedness satisfiability checker} then returns the satisfiability results to the \textsc{FOL$^*$ UNSAT proof analyzer}~\cite{feng-et-al-24c}, which analyzes the proof to compute a diagnosis. This diagnosis is subsequently transformed into a human readable format by the \textsc{well-formedness debugging support} component.

\subsection{Preliminary study}
\label{ssec:study}
We conducted an experiment with an ethicist, a lawyer, a philosopher, a psychologist, a safety analyst, a doctor, and two engineers to study effectiveness of \tool and identify the remaining challenges in eliciting normative requirements as \sleec{SLEEC} rules and analyzing their well-formedness using our first version of \sota~\cite{feng-et-al-24}. Participants were provided with a description of the system's capabilities and tasked with eliciting normative requirements, expressing them as \sleec{SLEEC} rules using the \sleec{SLEEC} DSL. Together, we then used \sota to assess the well-formedness of these rules. The tool provided feedback based on various well-formedness properties. In total, 233 rules were elicited, and 40 well-formedness issues were identified. Here, we summarize the main challenges and missing features that the stakeholders identified: 
\textbf{(1) No front-end support.} Stakeholders found it challenging and time-consuming to elicit requirements using the \sleec{SLEEC} DSL, because \sota did not provide syntax and semantic checking for such errors as  forgetting to define certain events, using incorrect data types, or having typos in identifiers.   As a result, stakeholders often had to guess the source of the issue, leading to frustration and delays in the elicitation process. 
\textbf{(ii) Excessive diagnostic information.} For situational conflicts, the diagnostic included a trace with event occurrences, but it also included all the values of the measures at each instant. However, only some of these values were useful for understanding the cause and resolving the conflict (see top of Fig.~\ref{fig:overExample}). 
\textbf{(iii) Mismatched diagnostics information.} For insufficient rules, the diagnostic provided a trace showing that the undesired behavior ("concern") could occur while respecting the rules (see top of Fig.~\ref{fig:overUnderExample}). However, it again included all the values of the measures, not just those relevant to the concern. Stakeholders also requested that the provided information focus only on which current \sleec{SLEEC} rules share similar concerns or system capabilities, excluding other unrelated information.

\subsection{Enhancing \tool}
To respond to user feedback, we have developed two new components, shown with a star ($\bigstar$).
The first, \textsc{IDE support}, offers both static (e.g., keywords or \sleec{SLEEC} rule skeletons) and dynamic (e.g., newly defined \sleec{events} or \sleec{measures}) code completion, alongside the code collapsing/expansion functionality. It also analyzes syntax and semantic errors in the elicited \sleec{SLEEC} rules. 
The second, 
 \textsc{Well-formedness Analysis}, enhances the diagnostic information for situational conflicts and insufficiency.


\begin{figure}
    \centering
    \includegraphics[scale=0.3]{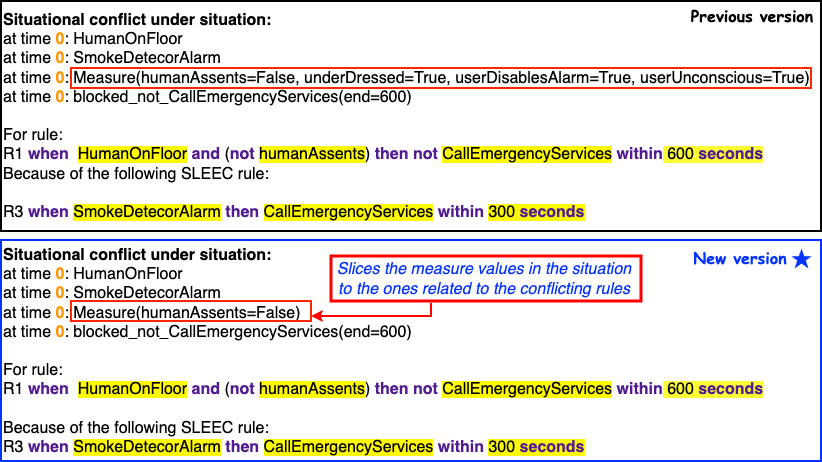}
    \caption{Comparison of previous and new \tool diagnostics for situational conflicting \sleec{SLEEC} rules. }
    \label{fig:overExample}
    \vspace{-0.15in}
\end{figure}

\boldparagraph{Excessive diagnostic information}
In \sleec{SLEEC DSL}, `\sleeckeyword{measures}' represent the ability to capture and provide information at any given time. In the context of situational conflicts, a diagnostic identifies a situation as a trace (e.g., top of Fig.~\ref{fig:overExample}) and the conflicting rules, and highlights the conflicting clauses. For each time point in the trace, a set of events occurs, and measure values are assigned. Since the trace includes all measure values for every time point, this can become overwhelming, especially in case studies involving a large number of measures. For instance, in the RESERVE repository~\cite{feng-et-al-24}, case studies included 25 measures in average. To address this, in \tool we simplified the situational diagnostic by showing, for each time point, only the values of the measures present in the trigger or defeater of the conflicting rules, as illustrated in the bottom of Fig.~\ref{fig:overExample}.

\begin{figure}
    \centering
    \includegraphics[scale=0.3]{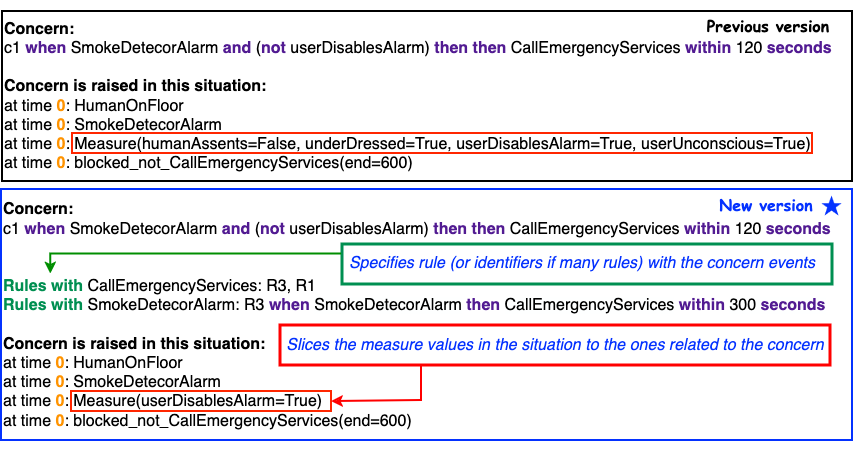}
    \caption{Comparison of previous and new \tool diagnostics for insufficient \sleec{SLEEC} rules.}
    \label{fig:overUnderExample}
    \vspace{-0.2in}
\end{figure}

\boldparagraph{Mismatched diagnostic information}
To 
ensure that a set of \sleec{SLEEC rules} is sufficient to prevent SLEEC harm, we define \emph{a concern} (a fact specifying undesirable behavior), such as \sleec{c1} in Tbl.~\ref{tab:sleecRules}, and verify whether the rules are sufficient to prevent it (see Sec.~\ref{sec:background}). If they are not, we produce an insufficiency diagnostic (e.g., top of Fig.~\ref{fig:overUnderExample}), which indicates the raised concern and provides an example of a situation (represented as a trace) where the concern occurs while all rules in the set are respected. 
This situation can be remedied by making existing rules stricter or adding additional rules.
However, the diagnostic does not provide the right information to facilitate this change: stakeholders need to parse the entire rule set and identify whether some rules could be applied. Additionally, similar to the situational conflict, the trace in the insufficiency diagnostic displays unnecessary  measure values at each time point. To address these issues, \tool augments the insufficiency diagnostic by indicating which rules contain the events related to the concern, and simplifies the situation by showing only the information relevant to the concern, as illustrated in the bottom of Fig.~\ref{fig:overUnderExample}.

\subsection{Implementation}
As shown in Fig.~\ref{fig:architecture-c}, \tool incorporates the \textsc{SLEEC to FOL* translator}~\cite{feng-et-al-23-b}, \textsc{FOL* wellformedness satisfiability checker}~\cite{feng-et-al-24}, the \tfol satisfiability checker \textsc{LEGOS}~\cite{feng-et-al-23-b}, the \textsc{FOL* UNSAT proof analyzer}~\cite{feng-et-al-24c}.  \tool  (available at \cite{DEMO-Artifact}) consists of 7,114 lines of code, primarily written in Java (6,821), with additional Backus-Naur Form grammar (202) and Flex (91). \tool IDE uses the IntelliJ Platform SDK.
\section{Preliminary Evaluation}
\label{sec:evaluation}
In this section, we evaluate the effectiveness of the new \tool component, \textsc{Well-formedness Analysis} (see Fig.~\ref{fig:architecture-c}). Specifically, we aim to answer \textbf{RQ:} How effective is the new diagnosis produced by \tool in helping  stakeholders understand and resolve the causes of situational conflicts, and concerns compared to the ones from the previous version? 
We conducted a preliminary controlled experiment with four practitioners from diverse backgrounds, including a philosopher (Ph), a software engineer (Se), a psychologist (Ps), and a doctor (Dc), and considered four case studies taken from the repository of normative requirements (before resolution)~\cite{feng-et-al-24}: 
(1) ALMI: a system assisting elderly; 
(2) ASPEN: an autonomous agent dedicated to forest protection; 
(3) BSN: a healthcare monitoring; 
(4) SafeAD: a driver attentiveness management system. 

To evaluate how the new diagnostic for situational conflicts and insufficiency helps stakeholders understand and resolve these issues, we divided them into two groups. Se \& Ph started by using the  previous version of \tool (\textit{v1-A}) on half of the case studies and then the version with the new diagnostic  (\textit{v2-B}) to perform the same task on the remaining case studies.  Stakeholders Dc \& Ps  began with the new tool (\textit{v2-A}) and then switched to the previous one (\textit{v1-B}) for the other case studies. 

For each situational conflict diagnostic, we recorded the name of the use case (case), the number of measures used (\#usedM) to understand and resolve the conflict out of the total number of measures present in the diagnostic, and the time taken for the resolution. For the v2's insufficiency diagnostic, we additionally recorded the number of rules used (\#helpedR) to understand and resolve the issue out of the total rules specified in the diagnostic. 

\begin{table}[t] 
    \caption{\small Usability of the enhanced diagnosis computed using \tool (\textit{v2}) compared to the previous version (\textit{v1}). 
    }
    \label{tab:results}
    \centering
    \scalebox{0.8}{
        \begin{tabular}{r c ||c c c|c c c c}
            \toprule
            \multirow{2}{*}{exp.} & \multirow{2}{*}{stak.} &  \multicolumn{3}{c}{s-conflict} & \multicolumn{4}{c}{insufficiency} \\
            \cline{3-6}\cline{7-9}
            & & case & \#usedM & time & case & \#usedM & \#helpedR & time\\ \toprule
            \emph{v1-A} & PS & ALMI & $3$/$15$ & $6.27$ & BSN  &  $0$/$31$ & - & $6.46$ \\
            & Dc & ALMI  & $3$/$15$ & $1.50$ & BSN  &  $0$/$31$ & - & $3.30$ \ \\
            \midrule
            \emph{v2-A}& \cellcolor{gray!10} SE & \cellcolor{gray!10} ALMI  &  \cellcolor{gray!10} $3$/$3$ & \cellcolor{gray!10} $2.42$ & \cellcolor{gray!10} BSN  &  \cellcolor{gray!10} $0$/$0$ & \cellcolor{gray!10} $1$/$3$ & \cellcolor{gray!10} $2.69$ \\
             & \cellcolor{gray!10} Ph & \cellcolor{gray!10} ALMI  &  \cellcolor{gray!10} $3$/$3$ & \cellcolor{gray!10} $2.36$ & \cellcolor{gray!10} BSN  &  \cellcolor{gray!10} $0$/$0$ & \cellcolor{gray!10} $2$/$3$ & \cellcolor{gray!10} $3.10$\\
             \midrule
           \emph{v1-B} & \cellcolor{gray!10} SE & \cellcolor{gray!10} SafeAD  &  \cellcolor{gray!10} $1$/$21$ & \cellcolor{gray!10} $3.51$ & \cellcolor{gray!10} ASPEN  &  \cellcolor{gray!10} $0$/$18$ & \cellcolor{gray!10} - & \cellcolor{gray!10} $5.82$ \\
             & \cellcolor{gray!10} Ph & \cellcolor{gray!10} SafeAD  &  \cellcolor{gray!10} $1$/$21$ & \cellcolor{gray!10} $3.39$ & \cellcolor{gray!10} ASPEN  &  \cellcolor{gray!10} $0$/$18$ & \cellcolor{gray!10} - & \cellcolor{gray!10} $4.26$ \\
             \midrule
            \emph{v2-B} & PS & SafeAD  & $1$/$3$ & 2.49 & ASPEN  &  $0$/$0$ & $1$/$9$ & $2.45$ \\
            & Dc & SafeAD  & $3$/$3$ & $1.15$ & ASPEN  &  $0$/$0$ & $4$/$9$ & $1.29$ \\
            \bottomrule
        \end{tabular}
    }
    \vspace{-0.15in}
\end{table}

\boldparagraph{Results}
The results are summarized in Tbl.~\ref{tab:results}.
Regardless of whether stakeholders started with the previous or the new version of \tool, all of them took less time to understand and resolve both the situational conflict and the insufficiency issues using diagnoses generated by the new \tool. This time difference is not only due to the relevance of the information provided in these diagnoses but also because of the IDE’s auto-completion feature in the new version. 
For situational conflict diagnoses, the new diagnosis filtered out more than 80\% of measure values as irrelevant information, thereby significantly helping the stakeholders understand the cause of the conflict. It also provided hints on whether a missing constraint on the measures should be added as a \sleec{trigger} or a \sleec{defeater} for the conflicting rules. Regarding insufficiency diagnoses, since the new version filtered out 100\% of the measure values as irrelevant information, it saved time but did not offer hints for resolution. However, the additional information in the new insufficiency diagnostic—- rules containing relevant events-—was helpful when only a few rules were suggested, as in the case of BSN, allowing stakeholders to determine whether they needed to edit existing rules or add new ones. When more than four rules were suggested, such as in ASPEN, it was less helpful as it became similar to manually parsing all of the rules.
{Thus, the answer to \textbf{RQ} is that \tool{}'s new diagnosis is effective in helping stakeholders understand and resolve situational conflicts. However, there is definitely room for improvement in the diagnosis for insufficiency issues, which we propose to address in future work.}



\section{Summary}
\label{sec:conclusion}
We presented \tool, a tool designed to support interdisciplinary stakeholders in specifying normative requirements as \sleec{SLEEC} rules and in analyzing and debugging their well-formedness.
Experiments conducted on four case studies with four interdisciplinary stakeholders demonstrated that new components in \tool were effective in aiding with understanding and resolving well-formedness issues.  Next steps of this project are extending \tool to 
suggest resolution patches for identified well-formedness issues.

\section*{Acknowledgment}
The authors  thank the four stakeholders for participating in our experiments; and  NSERC-CSE for funding this work.

\bibliographystyle{IEEEtran}
\bibliography{refs}

\end{document}